\documentclass[10pt,a4paper]{article}
\pdfoutput=1

\usepackage[T1]{fontenc}
\usepackage[utf8]{inputenc}

\usepackage{graphicx}
\usepackage{subcaption} % modern replacement for 'subfigure'

\usepackage{amsmath,amssymb,amsfonts}
\usepackage{mathrsfs,mathtools}
\usepackage{bm}  
\usepackage{slashed} % Dirac slash notation (\slashed{p})

\usepackage{xcolor}

\usepackage[a4paper, margin=1in]{geometry}
\usepackage{setspace}
\usepackage{cite}

\usepackage[
colorlinks=true,
linkcolor=blue,
citecolor=blue,
urlcolor=blue,
linktocpage=true
]{hyperref}

\definecolor{lgray}{gray}{0.6}

\newcommand{\nnb}{\nonumber}
\newcommand{\half}{\frac{1}{2}}
\newcommand{\tbt}{t_\beta}                     % tan(beta)
\newcommand{\tth}{t_\theta}                    % tan(theta)

\newcommand{\lamh}{\lambda_h}                   % Higgs self-coupling
\newcommand{\lams}{\lambda_s}                   % Scalar s self-coupling
\newcommand{\lamhs}{\lambda_{hs}}               % Portal coupling (h-s)
\newcommand{\lamhp}{\lambda_{h\phi}}             % Portal coupling (h-phi)
\newcommand{\lamsp}{\lambda_{s\phi}}             % Portal coupling (s-phi)

\title{\Large\bfseries 
	Self-interacting $O(N)$ scalar dark matter multiplet in a Higgs-portal model with a light scalar mediator
}

\author{
	U-Rae Kim$^{1}$\footnote{kim87@kma.ac.kr}, \quad
	Jungil Lee$^{2}$\footnote{jungil@korea.ac.kr}, \quad
	Soo-hyeon Nam$^{3}$\footnote{glvnsh@gmail.com}
	\\[4mm]
	\small $^{1}$\textit{Department of Physics, Korea Military Academy, Seoul 01805, Republic of Korea} \\
	\small $^{2}$\textit{Department of Physics, Korea University, Seoul 02841,  Republic of Korea} \\
	\small $^{3}$\textit{The Institute of Basic Science, Korea University, Seoul 02841,  Republic of Korea}
}

\date{}

\begin{document}

\maketitle

\begin{abstract}
	
We investigate self-interacting dark matter in an $O(N)$-symmetric real scalar multiplet model with Higgs-portal interactions involving the Standard Model Higgs and a real singlet scalar. After spontaneous breaking of the singlet $\mathbb{Z}_2$ symmetry, the singlet scalar mixes with the SM Higgs boson. We focus on the regime where the resulting light scalar serves as a mediator, leading to sizable and velocity-dependent self-interactions. We impose constraints from Higgs measurements, rare $B$-meson decays, the dark matter relic abundance, spin-independent direct-detection searches, and scalar coupling perturbativity. We find viable parameter regions in which destructive interference between scalar-mediated scattering amplitudes suppresses the direct-detection cross section while allowing self-interactions relevant to dwarf-galaxy scales. Among these viable parameter regions, a subset also satisfies electroweak vacuum stability up to the Planck scale. Finally, we show that future long-lived-particle searches can probe part of the viable parameter space, providing complementary tests of Higgs-portal self-interacting dark matter with a light scalar mediator.
	
\end{abstract}

%\pacs{PACS numbers:12.60.Fr,12.60.Cn,14.80.Cp}

%%%%%%%%%%%%%%%%%%%%%%%%%%%%%%%%%%%%%%%%%%%%%%%%%%%%%%%%%%%%%%%%
\section{Introduction} \label{sec:intro}
%%%%%%%%%%%%%%%%%%%%%%%%%%%%%%%%%%%%%%%%%%%%%%%%%%%%%%%%%%%%%%%%

The particle nature of dark matter (DM) remains one of the central open questions in particle physics and cosmology.
The standard cold dark matter (CDM) paradigm provides a successful description 
of the formation and evolution of large-scale structure, supported by a broad range of astrophysical and cosmological observations 
that establish the gravitational presence of DM over vastly different length scales.
At the same time, the assumption that DM is effectively collisionless has been challenged 
by several observations on galactic and sub-galactic scales, including the core--cusp problem, 
the diversity of galaxy rotation curves, and the too-big-to-fail problem.
Although the quantitative interpretation of these small-scale issues is subject to uncertainties associated 
with baryonic feedback and halo formation, they have motivated extensions of the collisionless-CDM paradigm 
in which DM possesses additional nongravitational interactions~\cite{Tulin:2017ara,Kaplinghat:2015aga}.

A particularly well-motivated possibility is self-interacting dark matter (SIDM), 
in which elastic DM self-scattering can alter the inner density profiles of halos 
while preserving the successful large-scale predictions of CDM~\cite{Spergel:1999mh}.
The self-interaction strength relevant for addressing small-scale structure is typically of order $0.1$--$10~\mathrm{cm^2/g}$, 
with the preferred range depending on the characteristic velocity of the astrophysical system~\cite{Tulin:2017ara,Adhikari:2022zvl}.
In particular, sizable self-interactions at the low velocities characteristic of dwarf galaxies must be compatible 
with constraints from larger and faster systems, such as galaxy clusters.
This tension generally favors a velocity-dependent scattering cross section, 
which can arise naturally when the self-interaction is mediated by a light particle 
rather than being described by a simple contact interaction.
Velocity-dependent SIDM is commonly associated with a mediator lighter than the DM particle.
The exchange of such a mediator generates a Yukawa potential between DM particles 
and can give rise to nonperturbative scattering, including resonant enhancements associated 
with near-threshold bound states~\cite{Loeb:2010gj,Tulin:2013teo}.
In generic light-mediator SIDM scenarios with weak-scale DM, 
mediator masses in the $\mathcal{O}(10\text{--}100)~\mathrm{MeV}$ range are 
often considered~\cite{Tulin:2013teo,Kaplinghat:2015aga}.
In this regime, a reliable calculation of the self-interaction cross section requires 
solving the nonrelativistic Schr\"odinger equation beyond the Born approximation.

From a particle-physics perspective, a light scalar mediator naturally motivates a dark sector 
coupled to the Standard Model (SM) through a portal interaction.
The Higgs portal provides one of the simplest renormalizable connections 
between a scalar dark sector and the SM~\cite{Silveira:1985rk,McDonald:1993ex,Burgess:2000yq,Arcadi:2019lka}.
When the mediator mixes with the SM Higgs boson, its mass and mixing angle are subject to collider, flavor, and low-energy constraints.
In particular, at sufficiently small masses, Higgs-mixed scalars are strongly constrained by rare meson decays 
and beam-dump experiments~\cite{Clarke:2013aya,Winkler:2018qyg,Fradette:2018hhl}.
These constraints motivate considering mediator masses above the very light regime, 
where nonperturbative self-scattering can still arise through resonant effects for DM in the electroweak-scale to sub-TeV mass range.

Beyond the mediator sector, scalar DM provides one of the simplest Higgs-portal DM scenarios.
The real scalar-singlet DM model has been extensively studied in the context of the DM relic abundance, direct detection, 
and Higgs phenomenology~\cite{Silveira:1985rk,McDonald:1993ex,Burgess:2000yq,GAMBIT:2017gge}.
A natural extension of this minimal framework is to promote the scalar DM field 
to an $N$-component multiplet transforming under an exact global $O(N)$ symmetry.
The unbroken symmetry stabilizes the DM multiplet and enforces mass degeneracy among its components 
without introducing an ad hoc discrete symmetry, while the multiplet structure leads to nontrivial dynamics in DM self-scattering.
In our previous work~\cite{Kim:2024eft}, we studied an $O(N)$-symmetric scalar DM model 
containing an additional real singlet scalar that acquires a vacuum expectation value and mixes with the SM Higgs boson.
We showed that the extended scalar sector can satisfy constraints from the DM relic abundance, direct detection, 
and Higgs measurements, while remaining perturbative and maintaining electroweak vacuum stability up to the Planck scale.
That analysis focused on the heavy-mediator regime, 
with the mediator mass required to exceed one half of the SM-like Higgs boson mass 
so that the exotic Higgs decay into a pair of mediators is kinematically forbidden.
In this regime, the mediator is too heavy to generate a sufficiently long-range Yukawa interaction at dwarf-galaxy velocities, 
and DM self-interactions were therefore not investigated.

In this work, we extend the $O(N)$-symmetric scalar DM framework to the light-mediator regime 
with mediator masses between $0.1~\mathrm{GeV}$ and $10~\mathrm{GeV}$, 
and investigate its viability as a velocity-dependent SIDM candidate.
This mass range allows nonperturbative DM self-scattering through the Yukawa potential 
while remaining accessible to complementary particle-physics and astrophysical constraints.
Since the DM multiplet consists of identical real scalars, the viscosity cross section is particularly suitable 
for characterizing the redistribution of momentum in astrophysical halos, 
as it accounts for the symmetry of identical-particle scattering by suppressing both forward and backward scattering.
We perform a comprehensive parameter scan in which the flavor-averaged viscosity cross section per unit DM mass is required 
to fall within the astrophysically favored range, while simultaneously imposing the observed DM relic abundance, 
spin-independent direct-detection limits, constraints from exotic Higgs decays and Higgs signal-strength measurements, 
and bounds from rare $B$-meson decays.
For viable SIDM configurations, we further examine the renormalization-group evolution of the scalar couplings 
to determine whether perturbativity and electroweak vacuum stability can be maintained up to the Planck scale.
Finally, we explore the future prospects for probing the remaining viable parameter space, showing that upcoming long-lived-particle searches can provide complementary tests for light scalar mediators in this framework.

The remainder of this paper is organized as follows.
In Sec.~\ref{sec:model}, we introduce the $O(N)$-symmetric scalar DM model and specify its scalar spectrum and interactions.
In Sec.~\ref{sec:constraints}, we summarize the theoretical, collider, flavor, relic-density, and direct-detection constraints.
Section~\ref{sec:selfint} presents the calculation of the DM self-interaction cross section, with particular emphasis on the flavor-averaged viscosity cross section in the nonperturbative regime.
Our numerical results, including the viable SIDM parameter space, the vacuum-stability analysis up to the Planck scale, and future prospects for long-lived-particle searches, are presented in Sec.~\ref{sec:numresults}.
Finally, Sec.~\ref{sec:conclusion} contains our conclusions.

%%%%%%%%%%%%%%%%%%%%%%%%%%%%%%%%%%%%%%%%%%%%%%%%%%%%%%%%%%%%%%%%
\section{The Model} \label{sec:model}
%%%%%%%%%%%%%%%%%%%%%%%%%%%%%%%%%%%%%%%%%%%%%%%%%%%%%%%%%%%%%%%%

We adopt a dark sector consisting of two real scalar fields $S$ and $\Phi$, which are SM gauge singlets
discussed first in our earlier study as the Type-II model of Ref.~\cite{Kim:2024eft}.
Since the theoretical framework has already been described in detail there, 
we only summarize the essential ingredients relevant for the present analysis and establish our notation.
The SM Higgs doublet $H$ is responsible for electroweak symmetry breaking (EWSB), the scalar mediator $S$ acquires a vacuum expectation value that spontaneously breaks the $\mathbb{Z}_2$ symmetry, 
and the scalar $\Phi$ is a DM candidate chosen to be the fundamental representation of a global O(N) group,
$\Phi=\left(\phi_1,\phi_2,\cdots,\phi_N\right)^T$.
%The exact $O(N)$ symmetry guarantees the stability of the DM particles and keeps all components degenerate in mass.
The extended Higgs sector Lagrangian is then given by
\begin{equation} 
\mathscr{L}_{\rm DM} = \left(D_\mu H\right)^{\dagger}D^\mu H 
+ \half\left(\partial^{\mu} S\right)^2 +\half(\partial_\mu\Phi^T)\partial^\mu\Phi - V(H, S, \Phi), 
\label{eq:DM_Lagrangian}
\end{equation}
where the scalar potential including the renormalizable Higgs portal and DM interactions is
\begin{align} 	
	V(H,S,\Phi) =&\, \mu_H^2 H^\dagger H + \lambda_h (H^\dagger H)^2 
	+ \half \mu_S^2 S^2 + \frac{1}{4} \lambda_s S^4 + \half \lambda_{hs} H^\dagger H S^2   \nnb \\
	& + \half \mu_\Phi^2 \Phi^T \Phi + \frac{1}{4}\lambda_\phi (\Phi^T \Phi)^2 
	+ \half \lambda_{h\phi} H^\dagger H \Phi^T \Phi + \frac{1}{4} \lambda_{s\phi} S^2 \Phi^T \Phi .
\label{eq:potential}	
\end{align}
%
%The electroweak symmetry breaking, scalar mass matrices, and their relations to the physical parameters 
%are identical to those derived in Ref.~\cite{Kim:2024eft} and are not repeated here.
The classically scale-invariant version of this model was studied in our previous work \cite{Jung:2019dog,Kim:2022sfc},
and we impose $Z_2$ symmetry on S for clear comparison with the scale-invariant case as similarly done in Ref.~\cite{Kim:2024eft}.
In these constructions, however, the additional theoretical relations imposed by classical conformal symmetry 
necessitate a relatively heavy mediator and a sizable mixing angle to satisfy the phenomenological constraints, 
thereby suppressing the DM self-interaction cross section.

The  scalar potential given in Eq.~\eqref{eq:potential} develops the nonzero vacuum expectation values (VEVs),
$\langle H^0 \rangle=v_h/\sqrt{2}$ and $\langle S \rangle = v_s$,
of the neutral component of the SM Higgs and the singlet scalar $S$, respectively,
while the DM scalar $\phi$ takes a vanishing VEV.
After the EWSB, 
the neutral scalar fields $h$ and $s$ defined by $H^0=(v_h+h)/\sqrt{2}$ and $S=v_s+s$ are mixed 
according to
\begin{equation}
\left( \begin{array}{c} h_1 \\[1pt] h_2 \end{array} \right) =
\left( \begin{array}{cc} \cos \theta &\ \sin \theta \\[1pt]
	-\sin \theta &\ \cos \theta \end{array} \right)
\left( \begin{array}{c} h \\[1pt] s \end{array} \right) ,
\label{eq:scalar_mixing}
\end{equation}
where $h_1$ is identified with the observed SM-like Higgs boson of mass $M_1=125~{\rm GeV}$,
while $h_2$ denotes the new lighter scalar mediator.
The mixing angle $\theta$ is expected to be very small (less than about 0.20) due
to the LEP constraints \cite{LEPWorkingGroupforHiggsbosonsearches:2003ing}.
After diagonalizing the mass matrix, 
we obtain the physical masses of the two scalar bosons ($h_1, h_2$) and the DM scalar $\phi$ as follows:
\begin{equation}
M^2_1 = \frac{2v_h^2(\lamh-\lams\tbt^2\tth^2)}{1-\tth^2}, \quad
M^2_2 = \frac{2v_h^2(\lams\tbt^2-\lamh\tth^2)}{1-\tth^2}, \quad
M_{\phi}^2 = \mu_\Phi^2 + \frac{v_h^2}{2}\left(\lambda_{h\phi} + \lambda_{s\phi}\tbt^2\right),
\end{equation}
where $\tth \equiv \tan\theta$ and $\tbt\, (\equiv \tan\beta) = v_s/v_h$.

The interactions relevant for the present analysis of DM self-interactions are 
the couplings of the scalar mediators to the DM particles,
\begin{equation}
	\mathscr{L} \supset -\frac12 g_{h_1\phi\phi} h_1\phi_i^2  -\frac12 g_{h_2\phi\phi} h_2\phi_i^2,
\end{equation}
with
\begin{align}
	g_{h_1\phi\phi} &= \lambda_{h\phi}v_h\cos\theta + 	\lambda_{s\phi}v_s\sin\theta, 
	\nnb \\
	g_{h_2\phi\phi} &= -\lambda_{h\phi}v_h\sin\theta + \lambda_{s\phi}v_s\cos\theta,
\end{align}
and with $\phi_i^2 = \Phi^T\Phi$.
The Higgs-singlet mixing also induces couplings of $h_2$ to the SM fermions and gauge bosons proportional to $\sin\theta$. Consequently, the model is constrained by Higgs signal strength measurements, exotic Higgs decays, rare meson decays, and beam-dump experiments. 
%The trilinear scalar coupling $g_{h_1h_2h_2}$, whose explicit expression is given in Ref.~\cite{OurPreviousPaper}, 
%determines the decay width of the exotic Higgs decay $h_1\to h_2h_2$ and 
%plays an important role in the light-mediator region considered in this work.
Given the fixed Higgs mass $M_1$ and $v_h \simeq 246$ GeV, 
there are seven independent new physics (NP) parameters:
$M_2, \tbt, \tth, M_\phi, \lambda_{h\phi}, \lambda_{s\phi}, \lambda_\phi $.
The dependencies of the model parameters are 
\begin{equation} 
v_s = v_h\tbt, \quad
\lamh = \frac{M_1^2 + M_2^2\tth^2}{2v_h^2(1+\tth^2)}, \quad
\lams = \frac{M_1^2\tth^2 + M_2^2}{2v_h^2\tbt^2(1+\tth^2)}, \quad
\lamhs = \frac{(M_1^2 - M_2^2)\tth}{v_h^2\tbt(1 + \tth^2)} .
\label{eq:couplings}
\end{equation}
It is clear from the above equation that $\tbt$ should not be very small 
because of the perturbativity of the couplings $\lambda_{hs}$ and $\lambda_{s}$.
By varying the above NP parameters, 
we investigate the phenomenology of this model by imposing collider constraints, 
DM relic abundance, direct detection limits, and DM self-interactions in the following sections. 
%Finally, we examine whether the surviving parameter space remains consistent 
%with vacuum stability up to high energy scales.

%%%%%%%%%%%%%%%%%%%%%%%%%%%%%%%%%%%%%%%%%%%%%%%%%%%%%%%%%%%%%%%%
\section{Constraints} \label{sec:constraints}
%%%%%%%%%%%%%%%%%%%%%%%%%%%%%%%%%%%%%%%%%%%%%%%%%%%%%%%%%%%%%%%%

%%%%%%%%%%%%%%%%%%%%%%%%%%%%%%%%%%%%%%%%%%%%%%%%%%%%%%%%%%%%%%%%
\subsection{Collider Constraints} \label{sec:collider}
 
In the light-mediator scenario considered in this work, if $M_2 \leq M_1/2$,
it is kinematically allowed for $h_1$ to decay into a pair of $h_2$.
Unlike the invisible Higgs decay scenario, the mediator $h_2$ does not decay into the DM particles 
because the mass hierarchy considered in this work satisfies $M_\phi \gg M_2$.
Therefore, the scalar mediator decays into the SM particles through its mixing with the Higgs boson, 
which contributes only to the total Higgs decay width.
The partial decay width of the process $h_1\rightarrow h_2h_2$ is given by
\begin{equation} 
	\Gamma(h_1\rightarrow h_2h_2) =
	\frac{g_{122}^2}{32\pi M_1}\sqrt{1-\frac{4M_2^2}{M_1^2}},
    \label{eq:width_h1h2h2}	
\end{equation}
where $g_{122}$ denotes the trilinear scalar coupling given as
\begin{equation}
g_{122} = \frac{2\tth(1+\tth\tbt)\left(\lamh (1-2\tth^2)+\lams (2-\tth^2)\tbt^2 \right)}
{(1-\tth^2)(1+\tth^2)^{\frac{3}{2}}\tbt}v_h .
\end{equation}
The mixing between the Higgs boson and the singlet scalar modifies the couplings of $h_1$ to the SM particles,
leading to a suppression of the SM decay widths by a factor of $\cos^2\theta$. 
Therefore, the total Higgs width in this model is
\begin{equation}	
	\Gamma_{h_1}^{\rm tot} 	= \Gamma_h^{\rm SM}\cos^2\theta + \Gamma(h_1\rightarrow h_2h_2),
\label{eq:higgs_total_width}	
\end{equation}
where $	\Gamma_h^{\rm SM}=4.07~{\rm MeV}$ is the SM prediction \cite{LHCHiggsCrossSectionWorkingGroup:2016ypw}.
The ATLAS and CMS collaborations have constrained the Higgs total width through direct measurements
as $\Gamma_\textrm{exp}= 3.7^{+1.9}_{-1.4}$ MeV \cite{ParticleDataGroup:2024cfk}. 
We impose the experimental upper limit at the $95\%$ confidence level.
%$\Gamma_{h_1}^{\rm tot}<\Gamma_h^{\rm exp}$
%where $\Gamma_h^{\rm exp}$ denotes the corresponding experimental bound.
%This constraint is particularly important in the present work because the light mediator required for sizeable DM %self-interactions opens the decay channel $h_1\rightarrow h_2h_2$, which was absent in our previous study with a heavy mediator.

The mixing between the new singlet scalar and the SM Higgs boson induces flavor-changing couplings 
of the scalar mediator through loop-level processes. 
If $h_2$ has a mass smaller than $M_B - M_K$,  it can be produced in rare $B$ meson decays,
and subsequently decays into the SM particles through its Higgs-like couplings 
proportional to the mixing angle $\theta$.
In particular, the transition $B^+\to K^+ h_2$ followed by $h_2\to\mu^+\mu^-$ provides stringent bounds 
on the Higgs-singlet mixing angle for $M_2\lesssim 4.8$ GeV. 
In the minimal Higgs-mixing scenario, the LHCb searches constrain the mixing angle down to 
$\sin\theta\sim10^{-4}$ in part of the GeV-scale scalar mass range \cite{Ovchynnikov:2023von}.

%%%%%%%%%%%%%%%%%%%%%%%%%%%%%%%%%%%%%%%%%%%%%%%%%%%%%%%%%%%%%%%%
\subsection{Dark Matter Constraints}  \label{sec:dmconstraints}

The DM relic abundance provides an important constraint on the parameter space of the model. 
Due to the exact $O(N)$ symmetry, all components of the scalar multiplet are stable and have identical masses. 
Therefore, the total relic abundance is determined by the combined contribution of the $N$ degenerate scalar DM particles.
The dark scalar particles $\phi_i$ undergo thermal freeze-out via pair-annihilation processes. 
In our relic-density analysis, we include all kinematically accessible $\phi_i\phi_i$ annihilation channels into SM particles 
through scalar mediators and into scalar final states. 
This comprehensive treatment ensures an accurate determination of the thermal relic abundance. 
The precise measurement of the DM relic abundance consequently imposes stringent constraints 
on the viable parameter space of the model. 
The current value of the DM density,  \(\Omega_{\rm DM} h^2 = 0.1200 \pm 0.0012\),  
has been determined by global fits to cosmological observations,  primarily based on the Planck satellite data,  
including the temperature and polarization anisotropies of the CMB,  
as well as its gravitational lensing measurements~\cite{Planck:2018vyg}.  
These precise measurements significantly limit the viable range of model parameters that can yield the correct thermal relic density.

In this work, the relic abundance is calculated using the numerical package micrOMEGAs \cite{Belanger:2018ccd,Alguero:2023zol}, 
which utilizes CalcHEP  \cite{Belyaev:2012qa} to compute the relevant annihilation cross sections,  
including all relevant annihilation channels. 
The Sommerfeld enhancement arising from long-range interactions mediated by the light scalar $h_2$ is incorporated 
through a custom implementation of the \texttt{improveCrossSection(...)} interface in micrOMEGAs \cite{Alguero:2023zol}. 
Specifically, for annihilation processes involving a pair of identical DM states, 
the annihilation cross section computed by micrOMEGAs is rescaled by a Hulth\'en-potential Sommerfeld factor $S(v_{\rm rel})$, 
evaluated as a function of the relative velocity $v_{\rm rel}$, 
the effective coupling $\alpha_2 = g_{h_2\phi\phi}^2/(16\pi M_{\phi}^2)$, and the mediator mass $M_2$. 
Annihilation processes involving two distinct DM components together are left unmodified. 
The Sommerfeld-corrected cross sections are then directly used in the thermal relic-density calculation. 
Since we assume that the $O(N)$ scalar particles constitute the entire DM component, 
we require the calculated relic abundance to be consistent with the observed value at the $3\sigma$ level.

The scalar DM particles can also be probed through their elastic scattering with nuclei 
via the exchange of the scalar mediators $h_1$ and $h_2$. 
Since both mediators couple to the DM particles and to the SM quarks through the Higgs portal interactions, 
the resulting SI scattering cross section provides an important constraint on the parameter space.
The effective DM--nucleon interaction receives contributions from both scalar mediators,
\begin{equation}
	\mathcal{L}_{\rm eff} = \frac{f_N}{2} \phi_i\phi_i \bar{\psi}_N\psi_N,
	\qquad
	f_N = \frac{m_N}{v_h}
	\left(\frac{g_{h_1\phi\phi}\cos\theta}{M_1^2} - \frac{g_{h_2\phi\phi}\sin\theta}{M_2^2}\right)
	f_N^{(s)},
	\label{eq:dm_nucleon_coupling}
\end{equation}
where $m_N$ is the nucleon mass, $\psi_N$ denotes the nucleon field, $f_N^{(s)}$ is the scalar nucleon form
factor, and the relative minus sign follows from the mixing convention of Eq.~\eqref{eq:scalar_mixing}. 
The corresponding SI cross section is
\begin{equation}
	\sigma_{\rm SI} = \frac{\mu_{\phi N}^2}{4\pi M_\phi^2}\,f_N^2,
	\qquad
	\mu_{\phi N} = \frac{M_\phi m_N}{M_\phi+m_N},
	\label{eq:sigma_SI}
\end{equation}
with $\mu_{\phi N}$ denoting the DM-nucleon reduced mass. The
interference between the $h_1$- and $h_2$-mediated contributions can lead to
a suppression of $\sigma_{\rm SI}$, giving rise to a blind-spot region
characterized approximately by
\begin{equation}
	\frac{g_{h_1\phi\phi}\cos\theta}{M_1^2}
	\simeq
	\frac{g_{h_2\phi\phi}\sin\theta}{M_2^2}.
\end{equation}
This effect is particularly relevant in the light mediator scenario, where
the $1/M_2^2$ dependence can significantly enhance the $h_2$-exchange
contribution.
For each parameter point, $\sigma_{\rm SI}$ is computed with micrOMEGAs and compared against recent direct detection bounds. 
Only points satisfying both the relic density and direct detection constraints are retained for the subsequent self-interaction analysis. Since all $O(N)$ components are degenerate and have identical interactions, 
the SI cross section is identical for each component; 
hence, we quote $\sigma_{\rm SI}$ for a single dark scalar component.

%%%%%%%%%%%%%%%%%%%%%%%%%%%%%%%%%%%%%%%%%%%%%%%%%%%%%%%%%%%%%%%%
\section{Dark Matter Self-interactions}\label{sec:selfint}
%%%%%%%%%%%%%%%%%%%%%%%%%%%%%%%%%%%%%%%%%%%%%%%%%%%%%%%%%%%%%%%%

%%%%%%%%%%%%%%%%%%%%%%%%%%%%%%%%%%%%%%%%%%%%%%%%%%%%%%%%%%%%%%%%
\subsection{Scattering channels and the observable}\label{subsec:channels}

For $N$ degenerate real scalars, the kinematically elastic $2\to2$ processes are classified into the three channels
\begin{equation}
	\text{I}:~\phi_i\phi_i\to\phi_i\phi_i,
	\qquad
	\text{II}:~\phi_i\phi_j\to\phi_i\phi_j~(i\neq j),
	\qquad
	\text{III}:~\phi_i\phi_i\to\phi_j\phi_j~(i\neq j),
	\label{eq:three_channels}
\end{equation}
whose multiplicities are $N$, $N(N-1)$, and $N(N-1)$, respectively. 
Although Type III changes the flavor labels of the outgoing particles, it is
kinematically elastic because all $\phi_i$ are exactly degenerate.
At tree level, the DM scalars interact through the quartic coupling $\lambda_\phi$ in
Eq.~(\ref{eq:potential}) and through $s$-, $t$-, and $u$-channel diagrams
mediated by $h_1$ and $h_2$. Owing to the $O(N)$ invariance of the couplings,
each channel involves a distinct subset of these diagrams, as summarized in
Table~\ref{tab:channels}.

\begin{table}[!hbt]
	\centering
	\caption{Tree-level diagrams contributing to the three channels of
		Eq.~(\ref{eq:three_channels}).}
	\label{tab:channels}
	\begin{tabular}{lcccc}
		\hline
		& contact & $s$ & $t$ & $u$ \\
		\hline
		Type I~~($\phi_i\phi_i\to\phi_i\phi_i$)
		& $\bullet$ & $\bullet$ & $\bullet$ & $\bullet$ \\
		Type II~~($\phi_i\phi_j\to\phi_i\phi_j$)
		& $\bullet$ & $-$       & $\bullet$ & $-$       \\
		Type III~($\phi_i\phi_i\to\phi_j\phi_j$)
		& $\bullet$ & $\bullet$ & $-$       & $-$       \\
		\hline
	\end{tabular}
\end{table}

The tree-level diagrams are divided into two groups according to whether their amplitudes depend on momentum. 
At leading order in the nonrelativistic expansion, $s = 4M_\phi^2 + \mathcal{O}(M_\phi^2 v_{\rm rel}^2)$, 
where $v_{\rm rel}$ denotes the relative velocity of the two incoming dark matter particles. 
Consequently, the contact and $s$-channel amplitudes can be treated as momentum-independent.
Their Fourier transforms are point-like interactions proportional to $\delta^{(3)}(\bm x)$ and act only in the $s$-wave. 
The $t$-channel amplitude depends on $(\bm p-\bm p')^2$ and the $u$-channel amplitude on $(\bm p+\bm p')^2$. 
Their Fourier transforms are long-range Yukawa potentials, which contribute to all partial waves. 
Type II contains a Yukawa potential and a contact interaction, 
Type I in addition the $s$-channel contribution and the $u$-channel exchange, 
and Type III the contact and $s$-channel contributions alone. 
Such a channel structure does not arise for single-component scalar DM ($N=1$), 
for which Types II and III are absent.

The $u$-channel diagram is obtained from the $t$-channel diagram by interchanging
the two final-state scalars and describes scattering into the angle
$\pi-\theta$. With the couplings of $h_1$ and $h_2$ diagonal in flavor, this
diagram exists only when all four scalars carry the same flavor, namely in Type I. 
A treatment based solely on the unsymmetrized $t$-channel amplitude is
therefore incomplete for Type I~\cite{Girmohanta:2022dog}. 
Section~\ref{subsec:matching} describes the treatment of the $u$-channel.

Since forward and backward scattering leave the DM phase-space distribution
nearly unchanged, the effect of self-interactions on halo structure is
characterized by an angular-weighted cross section rather than by
$\sigma_{\rm tot}$. For identical particles, scattering at $\theta$ and at
$\pi-\theta$ cannot be distinguished, and the appropriate measure is the
viscosity cross section~\cite{Tulin:2013teo,Tulin:2017ara}, defined as
\begin{equation}
	\sigma_V=\int d\Omega\,\sin^2\theta\,\frac{d\sigma}{d\Omega},
	\label{eq:sigmaV_def}
\end{equation}
which weights forward and backward scattering equally and emphasizes scattering
near $\theta=\pi/2$. We use $\sigma_V$ for all three channels, following
Refs.~\cite{Agrawal:2020lea,Girmohanta:2022dog}.

The relevant observable is $\sigma_V/M_\phi$ as a function of velocity.
Simulations of halo structure indicate that 
$\sigma/M_\phi\sim0.1$--$10~{\rm cm^2/g}$ on dwarf scales is sufficient 
to resolve the core-cusp and too-big-to-fail problems, 
while observations on Milky Way and cluster scales require $\sigma/M_\phi\lesssim0.1$--$1~{\rm cm^2/g}$, 
with characteristic velocities of approximately $10$, $200$, and $1000~{\rm km/s}$, respectively~\cite{Tulin:2013teo}.
 A velocity-independent cross section can satisfy both requirements 
 only within the narrow range $\sigma/M_\phi\sim0.1$--$1~{\rm cm^2/g}$.
With a light mediator, this restriction is relaxed because the Yukawa-mediated cross section becomes velocity-dependent. 
Away from resonances, the cross section decreases with increasing velocity 
once the momentum transfer becomes comparable to the mediator mass, 
while near a quasi-bound-state resonance it can exhibit the scaling $\sigma_V\propto v_{\rm rel}^{-2}$~\cite{Tulin:2013teo}. 
Consequently, the self-interaction cross section can remain large on dwarf scales 
while being sufficiently suppressed on cluster scales.

In the nonrelativistic regime, the elastic scattering of two DM scalars reduces
to a one-body problem with reduced mass $\mu=M_\phi/2$, and the relative momentum
is
\begin{equation}
	k=\mu\,v_{\rm rel}=\frac{M_\phi v_{\rm rel}}{2}.
	\label{eq:kdef}
\end{equation}
When the range of the mediator force, $1/M_2$, is comparable to or larger than the Bohr radius of the two-body system, 
a single exchange no longer suffices to describe the scattering. 
The repeated exchange of the mediator distorts the two-body wave function nonperturbatively. 
This long-range dynamics gives rise to Sommerfeld enhancement for short-distance reactions and, 
in elastic scattering, to resonant and antiresonant structures associated with near-threshold
quasi-bound states~\cite{Arkani-Hamed:2008hhe,Tulin:2013teo}.
These multiple-exchange contributions are resummed by mapping the scattering process onto 
an equivalent quantum-mechanical problem.

%%%%%%%%%%%%%%%%%%%%%%%%%%%%%%%%%%%%%%%%%%%%%%%%%%%%%%%%%%%%%%%%
\subsection{Matching onto a nonrelativistic potential problem}\label{subsec:matching}

We follow the matching procedure of Ref.~\cite{Agrawal:2020lea}. The
momentum-space potential $\tilde{V}(\bm{q})$ is obtained from the tree-level
amplitude $\mathcal{M}$ in the first Born approximation and reads
\begin{equation}
	\tilde V(\bm q)
	=
	-\frac{1}{4E_{p_f}E_{p_i}}\,\mathcal{M},
	\label{eq:matching}
\end{equation}
where $E_{p_i}, E_{p_f} \to M_\phi$ in the nonrelativistic limit. 
The Fourier transform of $\tilde V(\bm q)$ then yields the position-space potential $V(r)$.
At the Compton radius of the DM particle, $a=1/M_\phi$, the scattering problem is then
divided into two regions. For $r<a$ the collision is a short-distance,
high-energy process, for which the Sommerfeld enhancement does not develop and
the tree-level amplitude is adequate. Beyond $r=a$ the two scalars are
nonrelativistic and the multiple-exchange contributions must be resummed. The
Schr\"odinger equation is solved in the region $r>a$, while the region $r<a$
enters only through a boundary condition imposed at $r=a$.

Both scalar mass eigenstates contribute to the $t$-channel exchange, and the
long-range potential is
\begin{equation}
	V(r)
	=
	-\sum_{i=1}^{2}
	\alpha_i\,\frac{e^{-M_i r}}{r},
	\qquad
	\alpha_i
	\equiv
	\frac{g_{h_i\phi\phi}^{\,2}}{16\pi M_\phi^2},
	\label{eq:two_yukawa}
\end{equation}
where $g_{h_i\phi\phi}$ are the couplings defined in Sec.~\ref{sec:model}. The
potential in Eq.~(\ref{eq:two_yukawa}) is common to Types I and II, for which the
radial wave function $u_\ell(r)$ satisfies
\begin{equation}
	\left[
	\frac{d^2}{dr^2}
	+k^2
	-\frac{\ell(\ell+1)}{r^2}
	\right]u_\ell(r)
	=
	M_\phi\,V(r)\,u_\ell(r).
	\label{eq:radial}
\end{equation}
Type III has no $t$-channel exchange, so that $V^{\rm III}=0$, and it is treated
analytically in Sec.~\ref{subsec:phase}.

In terms of the dimensionless variable $x=M_2r$, the $h_2$-exchange term and
the kinetic term of Eq.~(\ref{eq:radial}) depend only on the two combinations
\begin{equation}
	\eta_2\equiv\frac{\alpha_2M_\phi}{M_2},
	\qquad
	\frac{M_\phi v_{\rm rel}}{M_2}=\frac{2k}{M_2}.
	\label{eq:eta}
\end{equation}
% which correspond to $b$ and $m_Xv/m_\phi$ of Ref.~\cite{Tulin:2013teo}. 
The parameter $\eta_2$ compares the force range with the Bohr radius
$a_B=2/(\alpha_2M_\phi)$ of the two-body system, while $M_\phi v_{\rm rel}/M_2$
compares it with the de~Broglie wavelength $1/k$.
The Born approximation applies for $\eta_2\ll1$, and the classical regime corresponds to
$M_\phi v_{\rm rel}/M_2\gg1$. In the remaining region, $\eta_2\gtrsim1$ and
$M_\phi v_{\rm rel}/M_2\lesssim1$, the potential supports quasi-bound states,
and the cross section exhibits resonances and antiresonances. Since no
analytic formula is available in this resonant regime, the Schr\"odinger
equation must be solved numerically~\cite{Tulin:2013teo}. 
Because $M_1\gg M_2$, the $h_1$ exchange is much shorter ranged than the $h_2$ exchange. 
In the parameter region considered here, its effect on the long-distance resonant structure is subleading, 
although it is retained in the potential and in the short-distance matching.

The contact and $s$-channel contributions are point interactions and cannot be
expressed as a potential of the form in Eq.~(\ref{eq:two_yukawa}). They are
instead imposed through a boundary condition at $r=a$ that encodes the
short-distance physics in the $K$-matrix element $K^a_\ell$. The boundary
condition reads
\begin{equation}
	\delta_\ell(a)=\arctan K^a_\ell ,
	\qquad
	K^a_\ell=\delta_{\ell0}\,K^{\rm c}_0-\frac{M_\phi}{k}\int_0^a s_\ell^2(kr)\,V(r)\,dr ,
	\label{eq:bc}
\end{equation}
where the integral is the part of the long-range potential inside $r=a$,
evaluated in the Born approximation. Being point-like, the contact and
$s$-channel contributions enter only through the $s$-wave term
\begin{equation}
	K^{\rm c}_0=-\frac{kM_\phi}{4\pi}\,T_X ,
	\qquad
	T_X = \begin{dcases}
		\frac{1}{2}\left[\frac{3\lambda_\phi}{2M_\phi^2}
		+\sum_{i=1}^{2}\alpha_i\,\frac{4\pi}{4M_\phi^2-M_i^2}\right],
		& (X=\mathrm{I})\\[1.5ex]
		\frac{\lambda_\phi}{2M_\phi^2},
		& (X=\mathrm{II})\\[1.5ex]
		\frac{\lambda_\phi}{2M_\phi^2}
		+\sum_{i=1}^{2}\alpha_i\,\frac{4\pi}{4M_\phi^2-M_i^2},
		& (X=\mathrm{III}).
	\end{dcases}
	\label{eq:Kc}
\end{equation}
Terms proportional to $\lambda_\phi$ arise from the
contact interaction, and those proportional to $1/(4M_\phi^2-M_i^2)$ from the
$s$-channel contribution. Although the potential of Eq.~(\ref{eq:two_yukawa})
applies to both Types I and II, their boundary conditions differ because the
$s$-channel term is absent in $T_{\rm II}$ (Table~\ref{tab:channels}).

In position space, the $u$-channel term of Type I is the $t$-channel potential
multiplied by the exchange operator $P_{\bm x}$, defined by
$P_{\bm x}\psi(\bm x)=\psi(-\bm x)$. For identical scalars the scattering
amplitude is symmetrized as $f(\theta)+f(\pi-\theta)$, and at leading order the
exchange term $f(\pi-\theta)$ reproduces the $u$-channel amplitude. We therefore
construct the Type I potential from the $t$-channel alone and restore the
$u$-channel by symmetrizing the asymptotic amplitude as in Eq.~(\ref{eq:fsym}).
Adding the $u$-channel term to the potential as well would count the $u$-channel
exchange twice, since $P_{\bm x}=1$ on the symmetric two-boson state. The contact
and $s$-channel contributions are isotropic and are doubled by the
symmetrization, so only half of each is retained in $T_{\rm I}$ of
Eq.~(\ref{eq:Kc}) to match the tree-level amplitude.

%%%%%%%%%%%%%%%%%%%%%%%%%%%%%%%%%%%%%%%%%%%%%%%%%%%%%%%%%%%%%%%%
\subsection{Phase shifts and cross sections}\label{subsec:phase}

The phase shifts of Types I and II are computed with the variable phase
method~\cite{Calogero_1967,Babikov_VPA}. In this method one introduces the phase
function $\delta_\ell(r)$, defined as the phase shift produced by the
interaction within radius $r$, so that the physical phase shift is the limit of
the phase function as $r\to\infty$. With the dimensionless variable $z=kr$, the
first-order nonlinear equation for the phase function reads
\begin{equation}
	\frac{d\delta_\ell}{dz}
	=
	-\frac{V(z/k)}{M_\phi\hat k^2}
	\Big[
	\cos\delta_\ell\,s_\ell(z)+\sin\delta_\ell\,c_\ell(z)
	\Big]^2,
	\label{eq:phase_eq}
\end{equation}
where $\hat k\equiv k/M_\phi$, and $s_\ell(z)=zj_\ell(z)$ and
$c_\ell(z)=-zy_\ell(z)$ are the Riccati-Bessel functions. The integration starts
at $z_0=ka$, where the initial value $\delta_\ell(z_0)$ is fixed by
Eq.~(\ref{eq:bc}). Contact and $s$-channel contributions enter the asymptotic
phase shift through this initial condition. Unlike the wave function, the phase
function does not grow exponentially inside the centrifugal barrier, so the
integration remains numerically stable~\cite{Calogero_1967}.

In Type II the two particles are distinguishable, all partial waves contribute,
and the viscosity cross section is
\begin{equation}
	\sigma_V^{\rm II}
	=
	\frac{4\pi}{k^2}
	\sum_{\ell=0}^{\infty}
	\frac{(\ell+1)(\ell+2)}{2\ell+3}
	\sin^2\!\left(\delta_{\ell+2}-\delta_\ell\right).
	\label{eq:sigmaV_II}
\end{equation}
For Type I the amplitude must be symmetrized. With
$P_\ell(-\cos\theta)=(-1)^\ell P_\ell(\cos\theta)$, the symmetrized amplitude is
\begin{equation}
	f_{\rm sym}(\theta)
	=
	f(\theta)+f(\pi-\theta)
	=
	\frac{1}{k}\sum_{\ell=0}^{\infty}
	(2\ell+1)\big[1+(-1)^\ell\big]\,
	e^{i\delta_\ell}\sin\delta_\ell\,P_\ell(\cos\theta).
	\label{eq:fsym}
\end{equation}
The factor $1+(-1)^\ell$ removes the odd partial waves and doubles the even
ones. With the phase-space factor $\tfrac12$ for identical particles, the
viscosity cross section is
\begin{equation}
	\sigma_V^{\rm I}
	=
	\frac{8\pi}{k^2}
	\sum_{\ell\,\rm even}
	\frac{(\ell+1)(\ell+2)}{2\ell+3}
	\sin^2\!\left(\delta_{\ell+2}-\delta_\ell\right).
	\label{eq:sigmaV_I}
\end{equation}

Type III has no long-range potential, so that $\delta_\ell=0$ for $\ell\ge1$ and
only the $s$-wave phase shift $\delta_0=\arctan K^a_0$ remains. In closed form,
the cross section reads
\begin{equation}
	\sigma_V^{\rm III}
	=
	\frac{2}{3}\,\frac{2\pi}{k^2}\,\sin^2\delta_0 ,
	\qquad
	\tan\delta_0
	=
	-\frac{kM_\phi}{4\pi}\,T_{\rm III},
	\label{eq:sigmaV_III}
\end{equation}
with $T_{\rm III}$ from Eq.~(\ref{eq:Kc}). The prefactor $2\pi/k^2$, rather than
$4\pi/k^2$, includes the symmetry factor $\tfrac12$ for the identical final state.
Equation~(\ref{eq:sigmaV_III}) incorporates the unitarization of the short-range
flavor-conversion amplitude through the $s$-wave phase shift. At low momentum $k$,
the phase shift $\delta_0$ is small, so that $\sin^2\delta_0 \simeq \tan^2\delta_0$
and Eq.~(\ref{eq:sigmaV_III}) reduces to the corresponding tree-level result.
This limit provides a check of the matching procedure.

The three channels are weighted by their multiplicities. Averaged
over initial flavors and summed over final flavors, the total viscosity cross
section is
\begin{equation}
	\sigma_V^{\rm tot}
	= \frac{1}{N}
	\Big[
	\sigma_V^{\rm I}
	+(N-1)\left(\sigma_V^{\rm II}+\sigma_V^{\rm III}\right)
	\Big].
	\label{eq:species_comb}
\end{equation}
By the $O(N)$ symmetry, all components are degenerate and equally abundant, so
$\sigma_V^{\rm tot}$ is the effective self-interaction cross section of the DM
fluid. When $N=1$, only the first term contributes and the single-component
result is recovered. For $N\ge2$ the weight of Type I is reduced to $1/N$, while
Types II and III each carry a weight $(N-1)/N$.

%%%%%%%%%%%%%%%%%%%%%%%%%%%%%%%%%%%%%%%%%%%%%%%%%%%%%%%%%%%%%%%%
\subsection{Numerical evaluation and astrophysical requirements}\label{subsec:numerics}

For each parameter point and relative velocity, we integrate the variable-phase
equation~(\ref{eq:phase_eq}) from $r=a$ to a sufficiently large radius at which
the potential is negligible and the phase shifts $\delta_\ell(r)$ have reached
their asymptotic values. The partial-wave sums in
Eqs.~(\ref{eq:sigmaV_II}) and~(\ref{eq:sigmaV_I}) are truncated once the
inclusion of additional partial waves changes the corresponding cross sections
by less than the target numerical precision. The three flavor channels are then
combined according to Eq.~(\ref{eq:species_comb}) to obtain the effective
viscosity cross section $\sigma_V^{\rm tot}$ of the $O(N)$-symmetric DM fluid.

We evaluate $\sigma_V^{\rm tot}/M_\phi$ at the characteristic relative
velocities of dwarf galaxies, Milky Way-sized halos, and galaxy clusters, and
require each parameter point to satisfy the astrophysical bounds specified in
Sec.~\ref{subsec:channels}. While the Milky Way- and cluster-scale conditions
impose upper limits on the self-interaction strength, the lower bound at
dwarf-galaxy velocities provides the most stringent restriction on the viable
parameter space. Parameter points satisfying these self-interaction criteria
are subsequently subjected to the collider and DM constraints described in
Sec.~\ref{sec:constraints}. The surviving parameter space, together with the
correlations among the DM mass $M_\phi$, mediator mass $M_2$, direct-detection
cross section, and $\sigma_V^{\rm tot}/M_\phi$, is presented in
Sec.~\ref{sec:directdetection}.

%%%%%%%%%%%%%%%%%%%%%%%%%%%%%%%%%%%%%%%%%%%%%%%%%%%%%%%%%%%%%%%%
\section{Numerical Results} \label{sec:numresults}
%%%%%%%%%%%%%%%%%%%%%%%%%%%%%%%%%%%%%%%%%%%%%%%%%%%%%%%%%%%%%%%%

In this section, we present the numerical results of the parameter scan, 
combining the theoretical, collider, and DM constraints discussed in Sec.~\ref{sec:constraints}
with the DM self-interaction condition developed in Sec.~\ref{sec:selfint}, 
in order to identify the viable region of the model parameter space.

%%%%%%%%%%%%%%%%%%%%%%%%%%%%%%%%%%%%%%%%%%%%%%%%%%%%%%%%%%%%%%%%
\subsection{Constraints on the parameter space} \label{sec:paramspace}

We perform a numerical scan over the independent model parameters introduced in Sec.~\ref{sec:model}. 
Although the initial scan is carried out over broader parameter ranges, after imposing the theoretical, collider, and DM constraints, 
the parameter region yielding phenomenologically relevant DM self-interactions is found to be concentrated 
in the ranges adopted for the numerical results:
\begin{align}
	\lambda_h &\in \left[0.11,\,0.135\right], \quad
	\lambda_{h\phi} \in \left[10^{-6},\, 0.4\right], \quad
	\lambda_{s\phi} \in \left[10^{-2},\, 0.4\right], \nnb \\
	\tan\theta &\in \left[10^{-10},\,10^{-4}\right], \quad
	M_2 \in \left[0.1,\,6\right]~{\rm GeV}, \quad
	M_\phi \in \left[63,\,800\right]~{\rm GeV}.
\end{align}
These ranges are used for the parameter-space plots presented in this section.

The mixing angle $\tan\theta$ is restricted to relatively small values. 
For $\tan\theta > 10^{-4}$, the constraints from rare $B$-meson decays and direct detection become increasingly restrictive, 
leaving no phenomenologically favored region with sizable DM self-interactions. 
The lower limit is chosen sufficiently small to cover the region 
in which the light mediator remains weakly coupled to the SM sector.

For the Higgs--DM portal coupling $\lambda_{h\phi}$ and the singlet--DM coupling $\lambda_{s\phi}$, 
extending these couplings beyond about $0.4$ does not lead to a phenomenologically favored region. 
Such large couplings are disfavored by the requirement of perturbativity up to the Planck scale 
and do not result in a significant enhancement of the self-interaction cross section. 
The lower limits are chosen to encompass the parameter region relevant for the relic abundance, 
direct detection, and self-interaction phenomenology.

For the light mediator mass, although we also scan the region up to $M_2=10~{\rm GeV}$, we find that the self-interaction cross section becomes too small to be phenomenologically relevant for $M_2\gtrsim6~{\rm GeV}$ after imposing the other constraints. 
This behavior is consistent with the mediator-mass dependence of the DM self-interaction discussed in Sec.~\ref{sec:selfint}.

For the DM mass, in our broader preliminary scan 
we find that the region below $M_\phi\approx63~{\rm GeV}$ does not yield phenomenologically relevant DM self-interactions 
after imposing the other constraints. 
We therefore restrict the scan to $M_\phi>63~{\rm GeV}$. 
This lower limit also lies slightly above the kinematic threshold for the invisible Higgs decays $h_1\to\phi_i\phi_i$, 
allowing us to avoid the additional constraint from the Higgs invisible decay width. 
On the other hand, extending the scan beyond $M_\phi=800~{\rm GeV}$ does not produce sizable values of $\sigma_V/M_\phi$.

We fix the scalar self-coupling and the singlet vacuum-expectation-value ratio 
to $\lambda_\phi=0.15$ and $\tbt=5$, respectively. 
Varying $\lambda_\phi$ has only a minor effect on the DM self-interaction cross section in the parameter region of interest, 
while values $\lambda_\phi\gtrsim 0.2$ are disfavored by perturbativity up to the Planck scale. 
Smaller values do not provide a significant improvement in electroweak vacuum stability. 
For $\tbt$, very small values can require relatively large singlet-sector couplings, 
such as $\lambda_{hs}$ and $\lambda_s$, potentially leading to perturbativity constraints as discussed in Sec.~\ref{sec:model}. 
Moreover, the self-interaction depends on combinations involving $\tbt$ and $\lambda_{s\phi}$, 
introducing a degeneracy between these parameters. 
We therefore fix $\tbt$ and vary $\lambda_{s\phi}$ explicitly.

The ranges above should be understood as the phenomenologically relevant region identified from the broader preliminary scan, 
rather than as strict prior boundaries of the model. 
In the following subsections, we use these ranges to investigate the surviving parameter space, 
the direct-detection and self-interaction phenomenology, 
and the vacuum stability and future search prospects of the light scalar mediator.

%%%%%%%%%%%%%%%%%%%%%%%%%%%%%%%%%%%%%%%%%%%%%%%%%%%%%%%%%%%%%%%%
\subsection{Direct detection and self-interaction} \label{sec:directdetection}

In our previous study of the $O(N)$ scalar DM model, we focused on the $N=2$ case. 
In the present analysis, we take $N=3$ as a representative case for the phenomenology. 
We have examined the cases $N=1,2,3,$ and $4$ and find that the resulting phenomenology is qualitatively similar, 
with no significant change in the overall structure of the allowed parameter space 
or the correlation between direct detection and self-interactions. 
Increasing $N$ generally enhances the renormalization-group running of the scalar couplings, 
making perturbativity up to high scales more restrictive. 
In addition, among the viable parameter points satisfying the self-interaction and other phenomenological constraints, 
the $N=3$ case provides more favorable benchmark points (BPs) for electroweak vacuum stability.

%%%%%%%%%%%%%%%%%%%%%%%%%%%%%%%%%%%%%%%%%%%%%%%%%%
\begin{figure}[!hbt]
	\centering%
	\includegraphics[width=12.0cm]{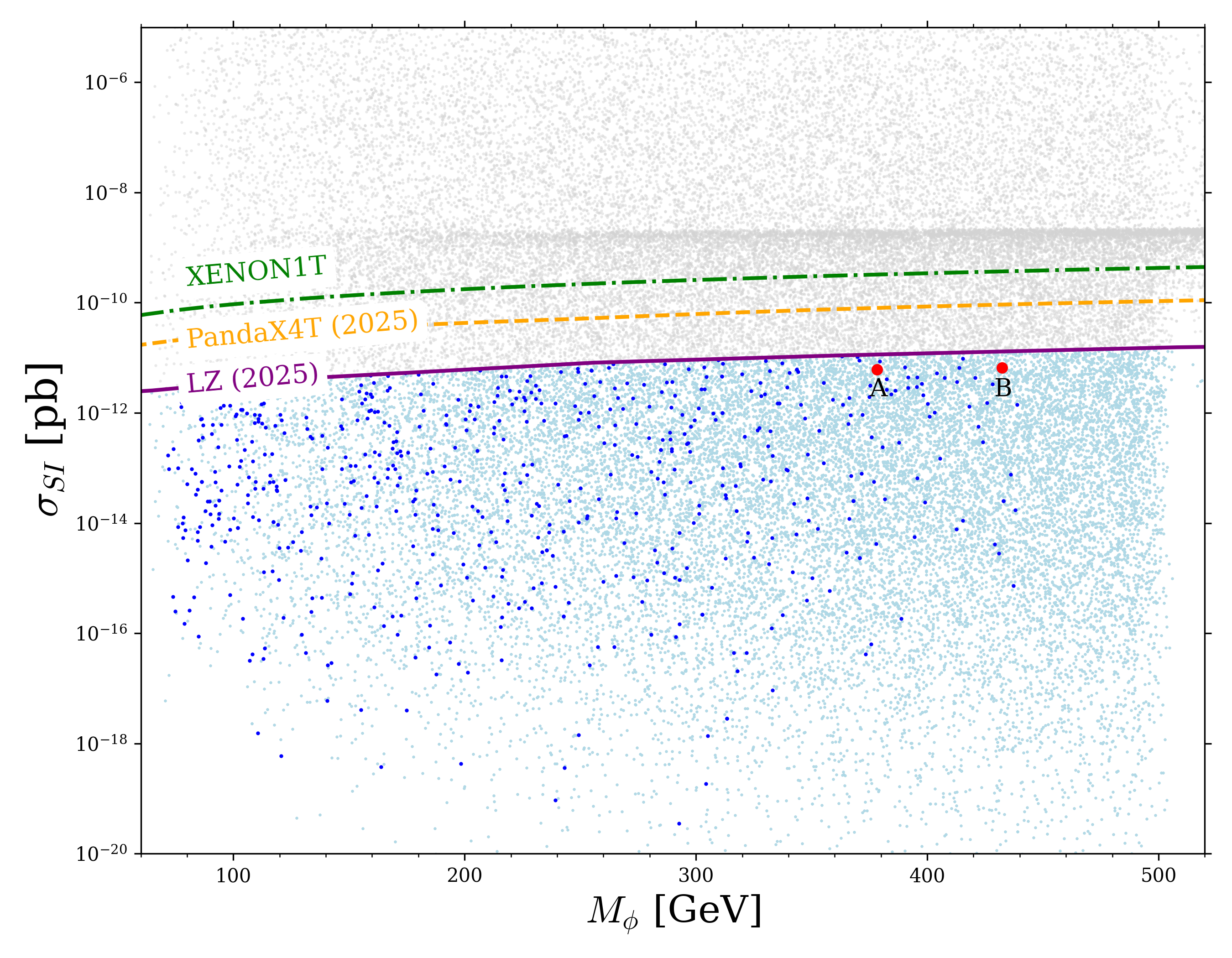} 
	\caption{Spin-independent DM-nucleon scattering cross section as a function of the DM mass $M_\phi$ 
		for parameter points with $N=3$ satisfying the theoretical, collider, and relic density constraints. 
		Also shown are the observed upper limits at the 90\% confidence level from XENON1T~\cite{XENON:2018voc}, 
		PandaX-4T~\cite{PandaX:2024qfu}, and LUX-ZEPLIN (LZ)~\cite{LZ:2024zvo}. 
		The parameter points excluded by the strongest LZ bound are shown in gray, 
		while the allowed points with $\sigma_V/M_\phi \geq 0.1~{\rm cm^2/g}$ and $\sigma_V/M_\phi<0.1~{\rm cm^2/g}$ 
		are shown in blue and light blue, respectively. 
		The two red dots mark the BPs selected for the subsequent analysis 
		of the running behavior of the scalar couplings and electroweak vacuum stability.
	}
	\label{fig:sigmaSI}
\end{figure}
%%%%%%%%%%%%%%%%%%%%%%%%%%%%%%%%%%%%%%%%%%%%%%%%%%

After imposing all theoretical, collider, and relic density constraints, 
we investigate the correlation between the DM direct detection signal and the self-interaction strength. 
The SI scattering cross section with nucleons is shown as a function of the DM mass in Fig.~\ref{fig:sigmaSI}. 
The surviving parameter points satisfy the Higgs total width constraint, rare $B$-meson decay constraints, 
and the observed relic abundance. 
We compare the resulting SI DM-nucleon scattering cross section 
with the upper limits at the 90\% confidence level from XENON1T~\cite{XENON:2018voc}, 
PandaX-4T~\cite{PandaX:2024qfu}, and LUX-ZEPLIN (LZ)~\cite{LZ:2024zvo}. 
The parameter points excluded by the strongest LZ bound are shown in gray, 
while the allowed points with $\sigma_V/M_\phi \geq 0.1~{\rm cm^2/g}$ and $\sigma_V/M_\phi<0.1~{\rm cm^2/g}$ are 
shown in blue and light blue, respectively. 
We display Fig.~\ref{fig:sigmaSI} only up to $M_\phi = 520~{\rm GeV}$, 
since the surviving parameter points below the LZ limit become sparse and
lie close to the exclusion bound for $M_\phi \gtrsim 500~{\rm GeV}$.
We find that the upper edge of this surviving region shifts toward lower DM masses as $N$ increases. 
In particular, for $N=3$, no points with $\sigma_V/M_\phi \geq 0.1~{\rm cm^2/g}$ are found above
$M_\phi \simeq 500~{\rm GeV}$.
The two red dots in the figure indicate the BPs selected for the subsequent analysis 
of the running behavior of the scalar couplings and electroweak vacuum stability.
Their numerical parameter sets are given by
\begin{align} 
	A:&\quad \tth = 2.35\times 10^{-5},\  M_2 = 1.28\ \textrm{GeV},\ M_\phi = 378\ \textrm{GeV},\
	    \lamhp = 0.196,\ \lamsp = 0.164, \nnb \\
	B:&\quad \tth = 1.96\times 10^{-6},\  M_2 = 0.36\ \textrm{GeV},\ M_\phi = 432\ \textrm{GeV},\
        \lamhp = 0.232,\ \lamsp = 0.182. 
\label{eq:benchmarks}        
\end{align}
The light mediator plays a crucial role in enhancing the self-interaction cross section. 
Since mediator exchange generates an attractive Yukawa potential, 
a smaller mediator mass $M_2$ increases the interaction range and can substantially enhance 
$\sigma_V/M_\phi$ relative to the heavy-mediator (contact-interaction) limit.

%%%%%%%%%%%%%%%%%%%%%%%%%%%%%%%%%%%%%%%%%%%%%%%%%%
\begin{figure}[!hbt]
	\centering%
	\includegraphics[width=12.0cm]{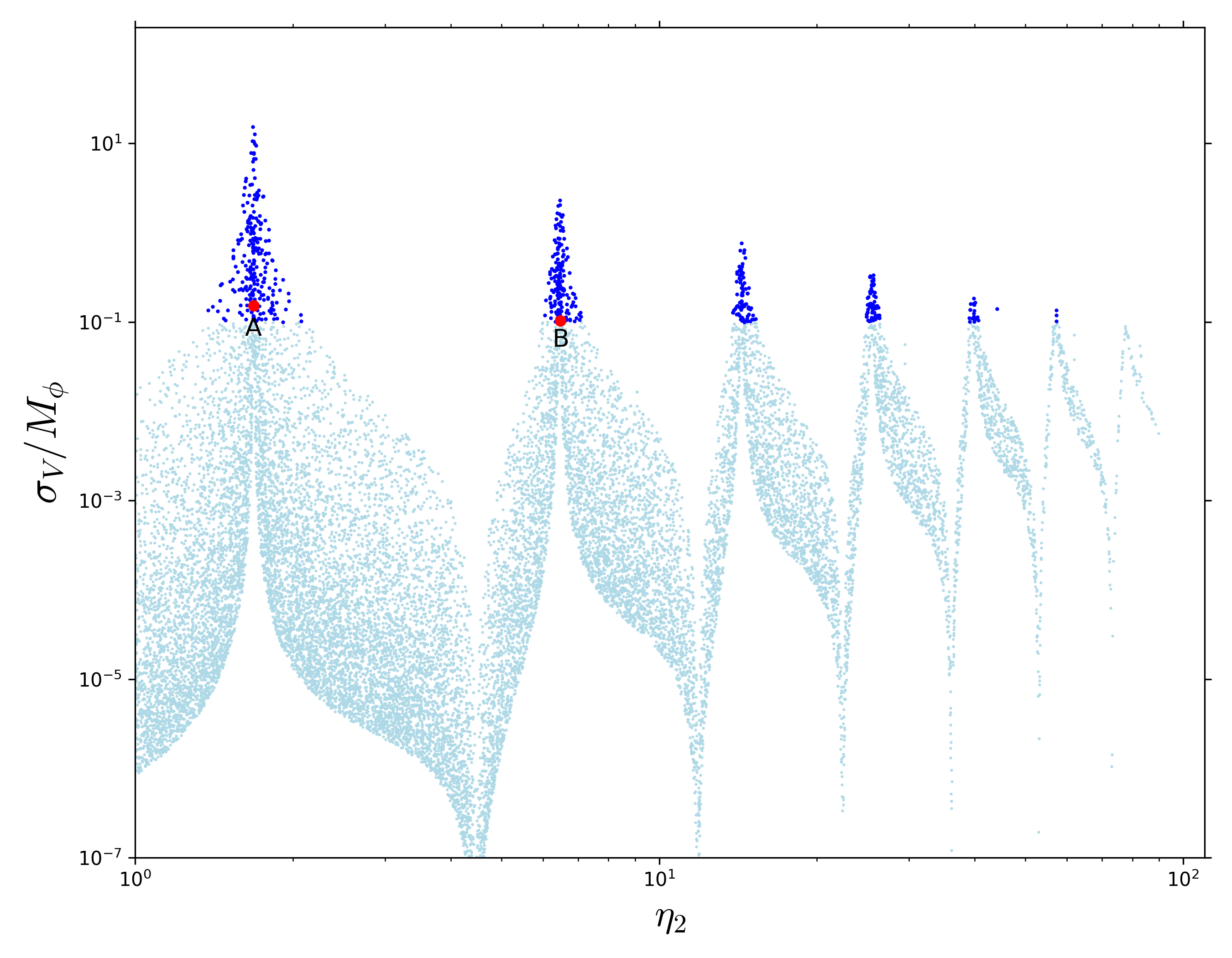} 
	\caption{The viscosity cross section per unit DM mass, $\sigma_V/M_\phi$, 
		as a function of $\eta_2$ for the parameter points allowed by the LZ direct-detection bound in the $N=3$ case. 
		The relative velocity is fixed to $v=10~{\rm km/s}$. 
		The points with $\sigma_V/M_\phi \geq 0.1~{\rm cm^2/g}$ and 
		$\sigma_V/M_\phi < 0.1~{\rm cm^2/g}$ are shown in blue and light blue, respectively. 
		The two red dots mark the BPs selected for the subsequent analysis of 
		the running behavior of the scalar couplings and electroweak vacuum stability.
	} 
	\label{fig:sigmav_eta}
\end{figure}
%%%%%%%%%%%%%%%%%%%%%%%%%%%%%%%%%%%%%%%%%%%%%%%%%%

In the resonant regime, the formation of (quasi-)bound states in the Yukawa potential 
can lead to a resonant enhancement of the self-interaction cross section. 
These resonant regions appear as narrow structures in the parameter space and 
cannot be captured by the perturbative Born approximation.
We evaluate the viscosity cross section using the nonperturbative treatment
described in Sec.~\ref{sec:selfint}, and present the resulting self-interaction strength,
$\sigma_V/M_\phi$, for the LZ-allowed points in the $N=3$ case, as a
function of $\eta_2 \equiv \alpha_2 M_\phi/M_2$, in
Fig.~\ref{fig:sigmav_eta}.
Following the choice adopted in Ref.~\cite{Tulin:2013teo} for dwarf-galaxy-scale halos, 
we take $v = 10~{\rm km/s}$ as a representative relative velocity in this calculation.
For the parameter ranges considered in the scan, the condition $M_\phi v/M_2 < 1$ is satisfied at this velocity, 
with the ratio ranging from $\sim 2\times10^{-3}$ to $\sim 0.2$. 
Thus, the variation with $\eta_2$ characterizes the transition from the perturbative Born regime to the nonperturbative regime, 
where nonperturbative effects become important for $\eta_2 \gtrsim 1$.
For a single attractive Yukawa potential, the parameter $\eta_i$ characterizes 
the strength of the potential relative to the mediator mass scale. 
In particular, the successive resonances are expected to occur parametrically at $\eta_2 \sim \kappa n^2$, 
where $n=1,2,\ldots$ labels the resonances and $\kappa$ is an order-one constant~\cite{Tulin:2013teo,Cassel:2009wt}.
This motivates the use of $\eta_2$ as the horizontal variable in Fig.~\ref{fig:sigmav_eta}. 
Although $\eta_2$ captures the dominant nonperturbative scale associated with the light mediator $h_2$, 
the precise resonance locations are obtained from the full two-mediator potential in Eq.~\eqref{eq:two_yukawa} 
using the numerical treatment described in Sec.~\ref{sec:selfint}.
For the two BPs indicated by the red dots in Fig.~\ref{fig:sigmav_eta}, 
the corresponding values of $\eta_2$ and
$\sigma_V/M_\phi$ are
\begin{align} \label{eq:benchmarks2}
	A:&\quad  \eta_2 = 1.68,\quad \sigma_V/M_\phi = 0.152, \nnb \\
	B:&\quad  \eta_2 = 6.47,\quad \sigma_V/M_\phi = 0.103.	
\end{align}
These two points lie at the first and second zero-energy $s$-wave resonances of the Yukawa potential. 
All points with $\sigma_V/M_\phi\geq0.1~{\rm cm^2/g}$ have $\eta_2>1$, 
indicating that the dwarf-scale requirement is met through the resonant enhancement of $s$-wave scattering.
A few points near the first resonance lie marginally above $\sigma_V/M_\phi = 10~{\rm cm^2/g}$ 
and are retained as blue points, 
taking account of the astrophysical uncertainty associated with this approximate upper limit.

%%%%%%%%%%%%%%%%%%%%%%%%%%%%%%%%%%%%%%%%%%%%%%%%%%
\begin{figure}[!hbt]
	\centering%
	\includegraphics[width=12cm]{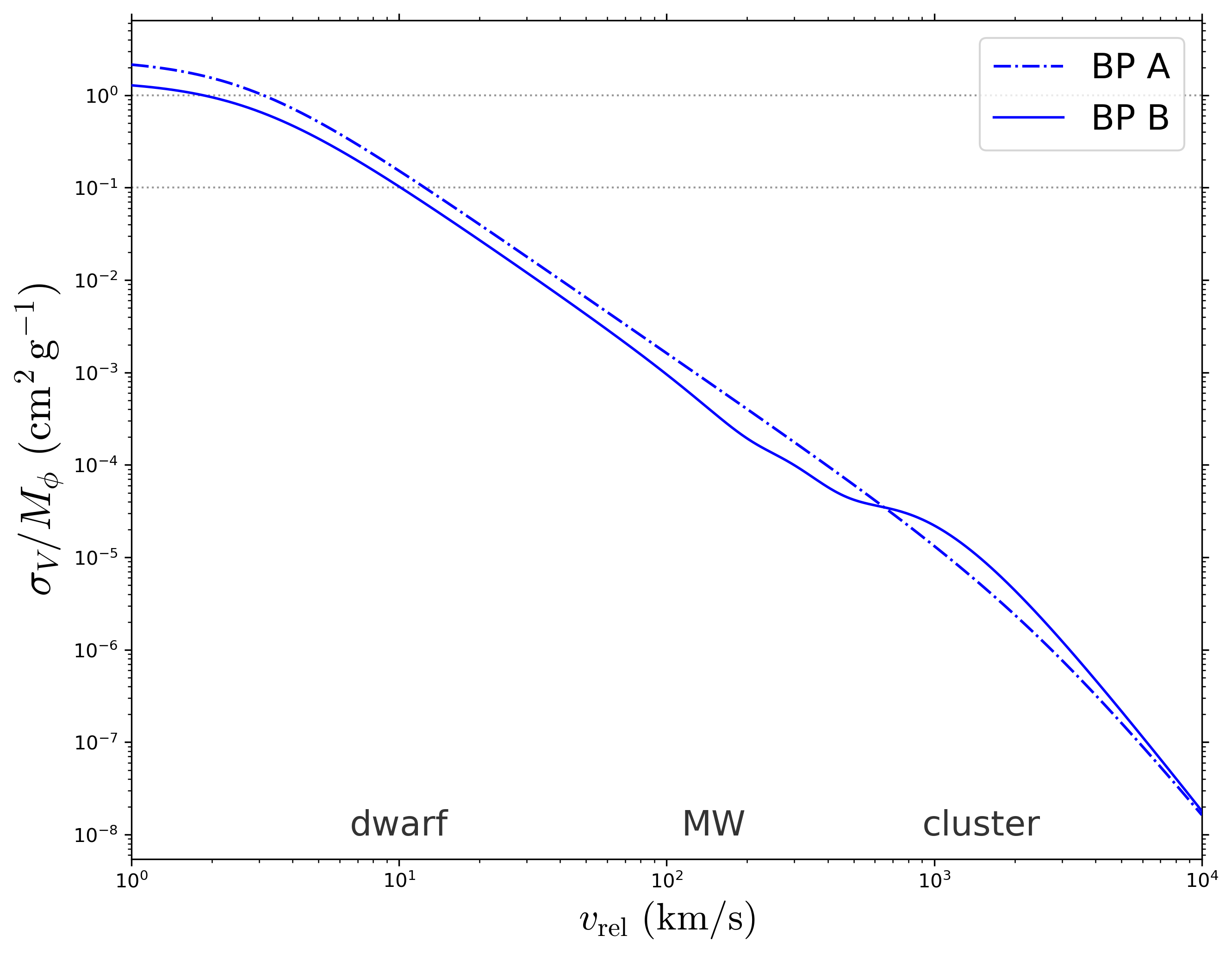}
     \caption{Velocity dependence of the viscosity self-interaction cross section per unit DM mass, 
     	$\sigma_V/M_\phi$, for BPs A (dash-dotted blue) and B (solid blue). 
     	The horizontal dotted lines indicate reference values of $\sigma_V/M_\phi = 0.1$ and $1~{\rm cm^2/g}$. 
     	The labels ``dwarf,'' ``MW,'' and ``cluster'' mark representative velocity scales 
     	for dwarf galaxies, the Milky Way, and galaxy clusters, respectively.
      }
	\label{fig:sigmaV_vel}
\end{figure}
%%%%%%%%%%%%%%%%%%%%%%%%%%%%%%%%%%%%%%%%%%%%%%%%%%

To illustrate the velocity dependence of the self-interaction for the BPs A and B, 
Fig.~\ref{fig:sigmaV_vel} displays $\sigma_V/M_\phi$ as a function of the relative velocity $v_{\rm rel}$.
Although both BPs feature similar values of $\alpha_2$
($\alpha_2 \simeq 5.7\times10^{-3}$ for A and
$5.3\times10^{-3}$ for B) and comparable DM masses
($M_\phi = 378$ and $432~{\rm GeV}$, respectively), their different values
of $\eta_2$ lead to somewhat different velocity dependences.
In the dwarf-galaxy velocity regime
($10 \lesssim v_{\rm rel} \lesssim 100~{\rm km/s}$), the self-interaction
cross section for both BPs decreases approximately as $v_{\rm rel}^{-2}$, from
$\sigma_V/M_\phi \simeq 0.10\text{--}0.15~{\rm cm^2/g}$ at
$v_{\rm rel}=10~{\rm km/s}$ to
$\sim 10^{-3}~{\rm cm^2/g}$ at $v_{\rm rel}=100~{\rm km/s}$.
This behavior is characteristic of nonperturbative scattering near a
resonance. At higher velocities, the self-interaction strength decreases
rapidly. Specifically, at $v_{\rm rel}=200$ and
$1{,}000~{\rm km/s}$, $\sigma_V/M_\phi$ is reduced to
$(2\text{--}4)\times10^{-4}$ and $(1\text{--}2)\times10^{-5}~{\rm cm^2/g}$,
respectively. Consequently, the BPs yield phenomenologically
relevant self-interactions at dwarf-galaxy velocities while becoming
strongly suppressed at Milky-Way and galaxy-cluster velocity scales.

The light mediator $h_2$ can induce sizable DM self-interactions, including nonperturbative resonant enhancements, 
while the SI direct-detection cross section can be suppressed by the destructive interference 
between the $h_1$- and $h_2$-exchange amplitudes. 
Consequently, parameter regions with $\sigma_V/M_\phi \gtrsim 0.1~{\rm cm^2/g}$ can remain compatible 
with the current LZ constraint. 
This interplay between direct detection and self-interaction phenomenology is a characteristic feature 
of the light-mediator $O(N)$ scalar DM model.
In the following subsection, we examine the electroweak vacuum stability of the BPs A and B,
selected from this viable region.

%%%%%%%%%%%%%%%%%%%%%%%%%%%%%%%%%%%%%%%%%%%%%%%%%%%%%%%%%%%%%%%%
\subsection{Vacuum stability of a viable SIDM benchmark} \label{sec:vacstability}

Having established the parameter space allowed by the collider,
relic-abundance, direct-detection, and self-interaction constraints in
Sec.~\ref{sec:directdetection}, we further investigate whether the viable
SIDM parameter space can also accommodate a stable electroweak vacuum.
To this end, we study the renormalization-group (RG) evolution of the
dimensionless scalar couplings
($\lambda_h$, $\lambda_s$, $\lambda_\phi$, $\lambda_{hs}$,
$\lambda_{h\phi}$, $\lambda_{s\phi}$) from the top-quark mass scale $M_t$
up to the Planck scale for BPs A and B with $N=3$.
We use the same RG equations as those employed in our previous work on this
model~\cite{Kim:2024eft}.
At each renormalization scale, we require all scalar couplings to remain
positive, which provides a sufficient condition for the stability of the
scalar potential for the BPs considered here. We also impose
the perturbativity condition
$\left|\lambda_i(\mu)\right|<4\pi$ up to the Planck scale.

In the SM, the running Higgs quartic coupling turns negative below
$M_{\rm Pl}$, mainly due to the negative contribution of the top Yukawa
coupling to its beta function, so that the electroweak vacuum is
metastable. In the present model, the scalar portal interactions modify the
coupled RG evolution of the scalar quartic couplings. In particular, a
sufficiently large $\lambda_{h\phi}$ at $M_t$ gives a positive contribution
to the running of $\lambda_h$, allowing it to remain positive up to
$M_{\rm Pl}$. This stabilization is not a generic feature of the viable
parameter space. Most of the allowed points in
Figs.~\ref{fig:sigmaSI}~and~\ref{fig:sigmav_eta} have $\lambda_{h\phi}$
values that are too small to compensate for the top-Yukawa contribution.
Their RG evolution consequently remains close to that in the SM, and the
electroweak vacuum is metastable.

%%%%%%%%%%%%%%%%%%%%%%%%%%%%%%%%%%%%%%%%%%%%%%%%%%
\begin{figure}[!hbt]
	\centering%
	\includegraphics[width=12cm]{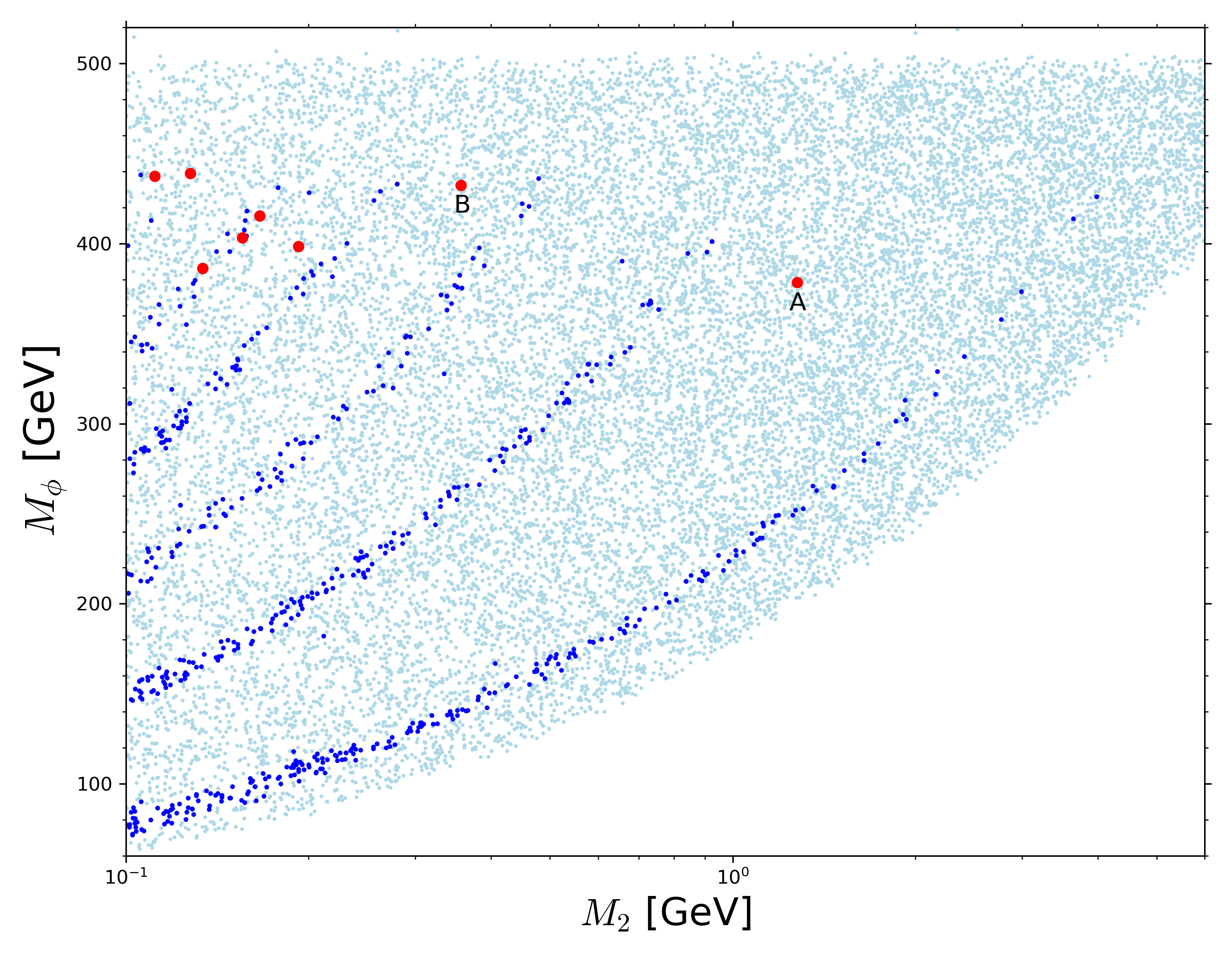}
	\caption{
		Distribution of the LZ-allowed parameter points in the $(M_2, M_\phi)$ plane for $N=3$. 
		The points with $\sigma_V/M_\phi \ge 0.1~{\rm cm^2/g}$ and $\sigma_V/M_\phi < 0.1~{\rm cm^2/g}$ 
		are shown in blue and light blue, respectively. The red points indicate the subset of blue points 
		that satisfy perturbativity and electroweak vacuum stability under renormalization-group evolution 
		for $\lambda_\phi = 0.15$ (corresponding to $\lambda_{h\phi} \gtrsim 0.18$ in the present scan).
	}
	\label{fig:M2_Mphi}
\end{figure}
%%%%%%%%%%%%%%%%%%%%%%%%%%%%%%%%%%%%%%%%%%%%%%%%%%

To examine how these requirements impact the viable model spectrum,
Fig.~\ref{fig:M2_Mphi} displays the distribution of parameter points in the
$(M_2, M_\phi)$ plane. The light and dark blue points have the same meaning
as in the previous figures, with the latter tracing a series of discrete
bands associated with near-threshold $h_2$-mediated resonances, as discussed
in Sec.~\ref{sec:selfint}. The red points denote the subset of dark blue
points that satisfy the RG requirements of perturbativity and electroweak
vacuum stability for $N=3$ and $\lambda_\phi=0.15$. In the present scan,
these points correspond to $\lambda_{h\phi}\gtrsim0.18$ and include
BPs A and B.

A larger $\lambda_{h\phi}$ enhances the Higgs-portal contribution to DM
annihilation and thus affects the relic-density calculation. Consequently,
points with $\lambda_{h\phi}\gtrsim0.18$ generally require different values
of $\lambda_{s\phi}$ to reproduce the observed relic density, and
hence a different ratio $M_\phi/M_2$ for the same resonant value of $\eta_2$.
As a result, the red points do not in general lie directly on the resonance bands traced by the
bulk of the dark blue points, even though they follow the same overall
correlation between $\sigma_V/M_\phi$ and $\eta_2$ shown in Fig.~\ref{fig:sigmav_eta}. 
In particular, BPs A and B lie on the first and second resonances in $\eta_2$ 
but away from the resonance bands in the $(M_2, M_\phi)$ plane.

%%%%%%%%%%%%%%%%%%%%%%%%%%%%%%%%%%%%%%%%%%%%%%%%%%
\begin{figure}[!ht]
	\centering%
	\includegraphics[width=7.6cm]{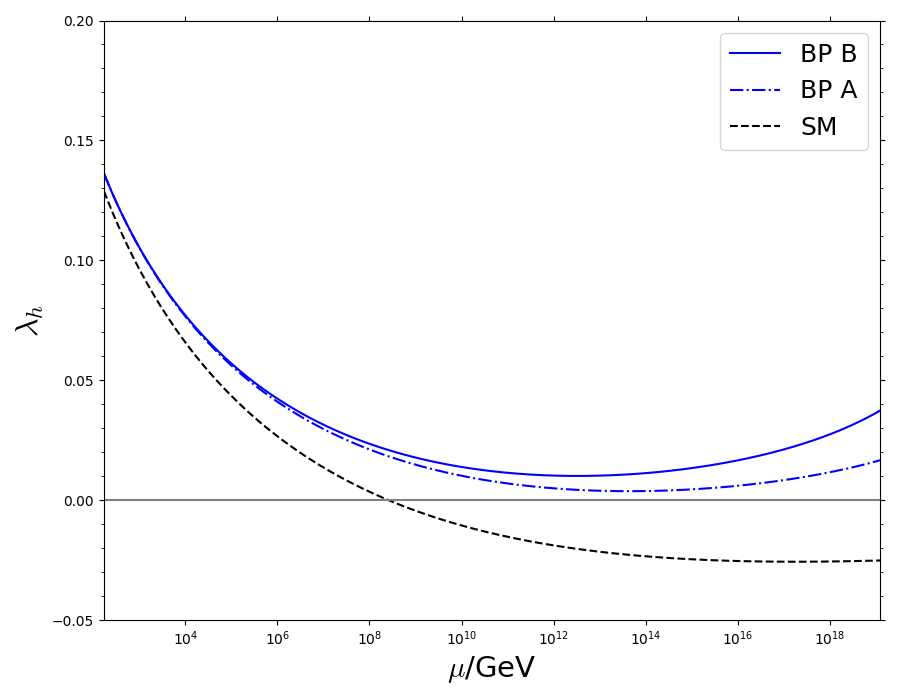} \quad
	\includegraphics[width=7.6cm]{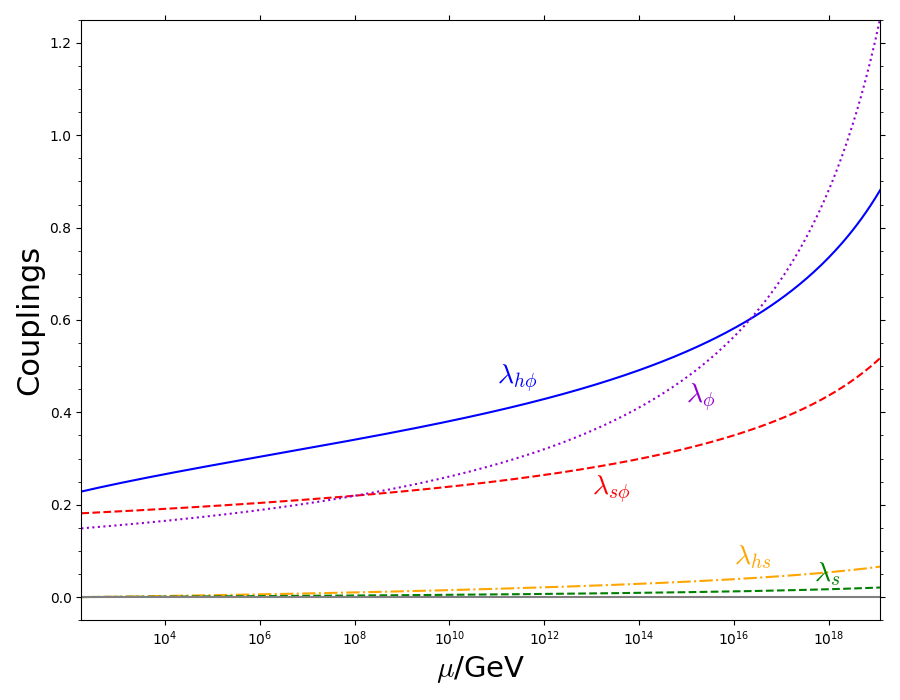} 
	\caption{Running of the dimensionless scalar couplings as functions of the renormalization scale $\mu$ for $N=3$.
		Left: RG evolution of the Higgs quartic coupling \(\lambda_h\) in the SM and for the BPs A and B.
		Right: RG evolution of the remaining scalar couplings for the BP B.
		For both BPs, the scalar vacua remain stable and all scalar couplings stay perturbative up to the Planck scale.
	}
	\label{fig:lambda_running}
\end{figure}
%%%%%%%%%%%%%%%%%%%%%%%%%%%%%%%%%%%%%%%%%%%%%%%%%%

Figure~\ref{fig:lambda_running} shows the resulting RG trajectories for
$N=3$. The left panel compares the running of $\lambda_h(\mu)$ in the SM
with those for BPs A and B. Unlike in the SM, $\lambda_h$
remains positive up to $M_{\rm Pl}$ for both BPs. The right
panel displays the running of the remaining scalar couplings for BP B.
The corresponding trajectories for BP A are
qualitatively similar, but lie closer to one another and overlap
substantially in the same panel. We therefore show only BP B
in the right panel to avoid overcrowding. For both BPs, every
scalar coupling remains positive throughout the RG evolution, and none
exceeds $\lambda_i\simeq1.3$ at $M_{\rm Pl}$, well below the perturbativity
bound.

These results demonstrate that a subset of the viable SIDM parameter space
can simultaneously satisfy the phenomenological constraints discussed in
Sec.~\ref{sec:directdetection} and the theoretical requirements of
perturbativity and electroweak vacuum stability up to the Planck scale.
BPs A and B therefore illustrate a possibility realized within
the viable SIDM parameter space, rather than a property shared by most
allowed points.

%%%%%%%%%%%%%%%%%%%%%%%%%%%%%%%%%%%%%%%%%%%%%%%%%%%%%%%%%%%%%%%%
\subsection{Future prospects for scalar mediator searches} \label{sec:prospects}

The viable parameter space identified in Sec.~\ref{sec:paramspace} extends to very small Higgs--scalar mixing angles, 
as a consequence of the $h_1$--$h_2$ interference cancellation discussed in Sec.~\ref{sec:directdetection}, 
which keeps the SI direct-detection cross section below current limits.
Since the couplings of the light mediator $h_2$ to SM particles are induced entirely by this mixing, 
its visible decay width is suppressed by $\sin^2\theta$, and its decay length grows correspondingly for smaller mixing angles.  
The light mediator can therefore be long-lived in part of the allowed parameter space, 
motivating a comparison with intensity-frontier, forward, and displaced-vertex searches for long-lived particles (LLPs).

%%%%%%%%%%%%%%%%%%%%%%%%%%%%%%%%%%%%%%%%%%%%%%%%%%
\begin{figure}[!hbt]
	\centering%
	\includegraphics[width=12.0cm]{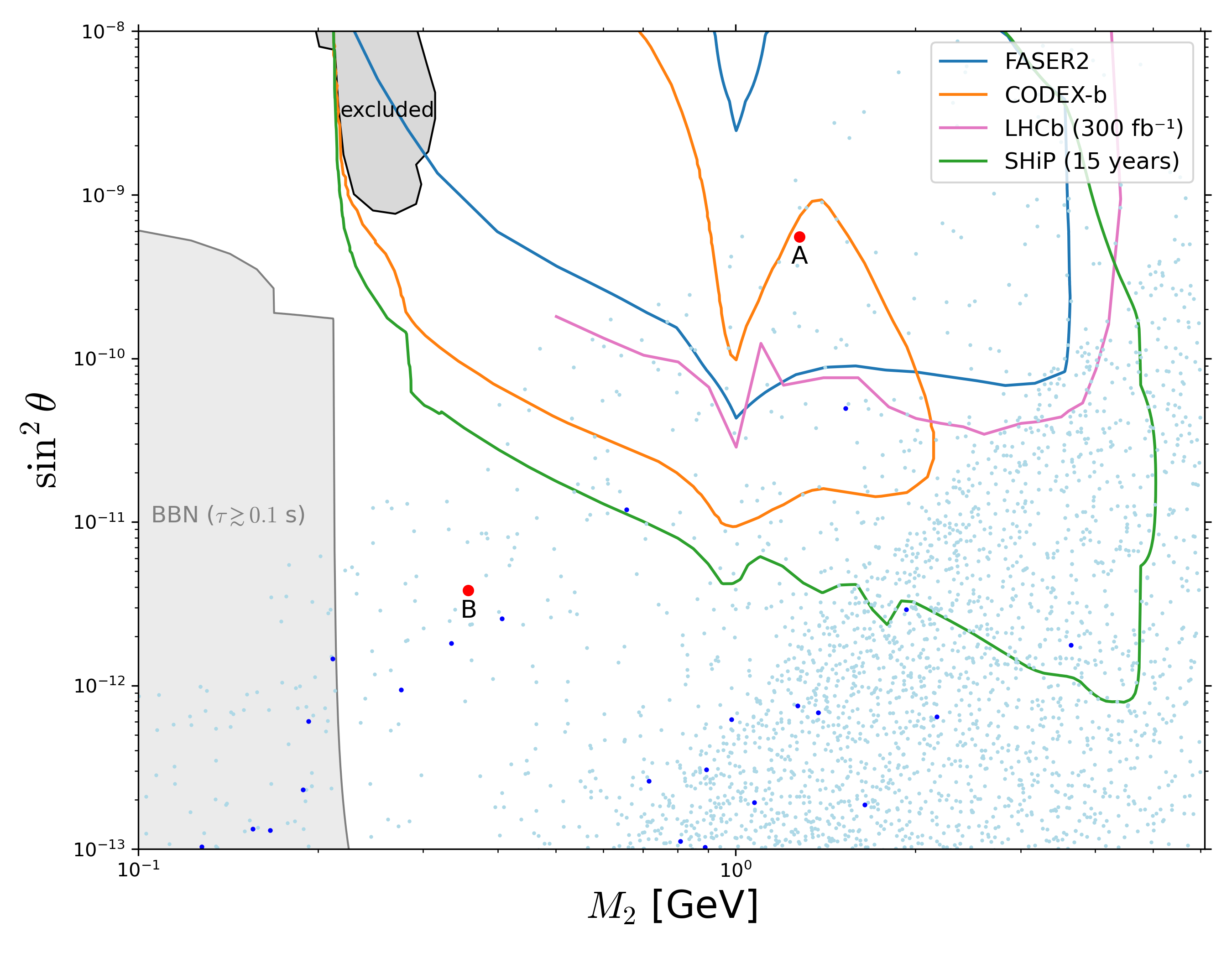} 
	\caption{Projected sensitivities in the $(M_2,\sin^2\theta)$ plane from SHiP 
		after 15 years of operation~\cite{SHiP:2025sr370},
		FASER2~\cite{Feng:2022inv}, CODEX-b~\cite{CODEX-b:2025rck}, 
		and the LHCb tracker--trigger upgrade with $300~\mathrm{fb}^{-1}$ \cite{Alimena:2025fips}, 
		together with the viable parameter points obtained in this work. 
		The upper gray region is excluded by neutrino-beam and beam-dump searches~\cite{PBC:2025sny},
		while the lower gray region is excluded by BBN and SN1987A cooling considerations~\cite{Fradette:2017sdd,Dev:2020eam}. 
		The blue and light-blue viable points correspond to $\sigma_V/M_\phi$ above and 
		below the dwarf-scale target value $0.1~\mathrm{cm^2/g}$, respectively, 
		and the two red dots mark the BPs.
	} 
	\label{fig:prospects}
\end{figure}
%%%%%%%%%%%%%%%%%%%%%%%%%%%%%%%%%%%%%%%%%%%%%%%%%%
 
Although $h_1\to h_2h_2$ is kinematically open throughout the parameter space considered here, 
we find $\mathrm{BR}(h_1\to h_2h_2)\lesssim10^{-11}$ for all viable points after imposing the constraints discussed above.  
Exotic Higgs decays therefore make a negligible contribution to $h_2$ production.  
The viable points can consequently be compared directly with the minimal scalar-portal benchmark model 
with $\mathrm{BR}(H\to SS)=0$ (BC4), rather than with the BC5 benchmark 
involving a non-vanishing exotic Higgs decay branching fraction~\cite{Beacham:2019nyx}.  
The future sensitivity curves shown in Fig.~\ref{fig:prospects} are obtained for this BC4 case: 
SHiP~\cite{SHiP:2025sr370}, FASER2~\cite{Feng:2022inv}, CODEX-b~\cite{CODEX-b:2025rck}, 
and the LHCb tracker--trigger upgrade with $300~\mathrm{fb}^{-1}$~\cite{Alimena:2025fips}. 
In particular, the SHiP curve corresponds to 15 years of running, or $6\times10^{20}$ protons on target, 
as reported in the BDF/SHiP Annual Report 2025~\cite{SHiP:2025sr370}.

Figure~\ref{fig:prospects} shows the viable points from Fig.~\ref{fig:sigmav_eta} in the $(M_2,\sin^2\theta)$ plane, 
together with present exclusions and the future sensitivities noted above. 
We focus on the mass and mixing-angle window over which the selected reference sensitivity curves are shown. 
Most viable points have $\sin^2\theta$ far below the lower edge of the displayed range and
therefore lie outside the plotted region; 
they are not excluded for this reason. 
Points are colored as in Figs.~\ref{fig:sigmaSI} and~\ref{fig:sigmav_eta}, 
with BPs A and B marked separately. 
Within the displayed range, the upper excluded region arises from searches for $K\to\pi h_2$ with $h_2\to\ell^+\ell^-$ 
($\ell=e,\mu$) at neutrino-beam experiments and from reinterpretations of beam-dump data~\cite{PBC:2025sny}.
The lower gray region combines the BBN bound on sufficiently long-lived scalars, 
with a characteristic lifetime scale of $\tau_{h_2}\gtrsim 0.1~\mathrm{s}$~\cite{Fradette:2017sdd}, 
and the SN1987A energy-loss bound~\cite{Dev:2020eam}.

As shown in Fig.~\ref{fig:prospects}, future LLP experiments can probe part of the viable SIDM parameter space 
at comparatively larger values of $\sin^2\theta$.  
A number of viable points lie within the projected reaches of FASER2, CODEX-b, 
the LHCb tracker--trigger upgrade, and SHiP.  
Most of these points lie below the dwarf-scale target value $\sigma_V/M_\phi = 0.1~\mathrm{cm^2/g}$, 
while a few lie above it.
Benchmark point A is among these, lying within all four projected sensitivity regions 
and close to the CODEX-b boundary in particular.
Benchmark point B, together with the majority of viable points at smaller mixing angles, 
remains beyond all four projected sensitivities shown in the figure.  
Searches for displaced decays of Higgs-mixed scalars thus provide a complementary probe 
of part of the light-mediator SIDM parameter space, 
whereas an ultraweak-mixing region, exemplified by point B, 
remains beyond the projected reach of the experiments considered here.

%%%%%%%%%%%%%%%%%%%%%%%%%%%%%%%%%%%%%%%%%%%%%%%%%%%%%%%%%%%%%%%%
\section{Conclusion}  \label{sec:conclusion}
%%%%%%%%%%%%%%%%%%%%%%%%%%%%%%%%%%%%%%%%%%%%%%%%%%%%%%%%%%%%%%%%

We have investigated SIDM in an $O(N)$-symmetric real scalar multiplet model with a real singlet scalar mediator. 
The singlet acquires a vacuum expectation value through the spontaneous breaking of its $\mathbb{Z}_2$ symmetry 
and mixes with the SM Higgs boson, giving rise to the SM-like Higgs boson $h_1$ and a lighter scalar mediator $h_2$. 
We have focused on the light-mediator region $0.1~{\rm GeV}<M_2<10~{\rm GeV}$, 
extending the analysis of the Type-II model of Ref.~\cite{Kim:2024eft} to a regime phenomenologically distinct 
from the heavier-mediator case studied previously, 
where the decay $h_1\to h_2h_2$ is kinematically open.

We imposed collider, cosmological, and direct-detection constraints on the model. 
The collider constraints include the Higgs-width bound on the exotic decay $h_1\to h_2h_2$ 
and LHCb searches for rare $B$-meson decays, 
which constrain the scalar mixing angle in part of the GeV-scale mediator-mass range. 
We further required $\Phi$ to reproduce the observed DM relic abundance 
and applied current spin-independent direct-detection bounds.
Although the $h_1\to h_2h_2$ decay is open throughout the light-mediator region, 
its branching fraction is below $10^{-11}$ for all viable points.
Exotic Higgs decays therefore have a negligible impact on the phenomenology of the surviving parameter space.

A characteristic feature of the viable region is the cancellation between the $h_1$- and $h_2$-mediated contributions 
to the spin-independent DM--nucleon scattering amplitude. 
This cancellation allows the direct-detection cross section to evade current limits, 
particularly the strongest bound from LZ, while retaining sizable DM self-interactions. 
In particular, viable points with $\sigma_V/M_\phi \geq 0.1~{\rm cm^2/g}$, relevant to dwarf-galaxy scales, 
are obtained in the light-mediator regime. 
For BPs A and B, the self-interaction cross section decreases approximately as $v_{\rm rel}^{-2}$ 
over the dwarf-galaxy velocity range and becomes strongly suppressed at Milky-Way and galaxy-cluster velocity scales. 
This illustrates the velocity-dependent SIDM phenomenology realized in the viable parameter space.

The renormalization-group evolution of the scalar couplings further shows 
that electroweak vacuum stability up to the Planck scale is not a generic feature of the viable SIDM parameter space. 
Most viable points have $\lambda_{h\phi}$ values that are too small to prevent the Higgs quartic coupling from becoming negative, 
and consequently exhibit an electroweak vacuum metastable in a manner similar to that of the SM. 
In contrast, a smaller subset with sufficiently large $\lambda_{h\phi}$ satisfies both perturbativity 
and electroweak vacuum stability up to the Planck scale. 
BPs A and B provide explicit examples of this stable subset.

Finally, we compared the viable parameter space with projected long-lived-particle sensitivities 
at SHiP, FASER2, CODEX-b, and the LHCb tracker--trigger upgrade. 
Since the viable points have a negligible exotic Higgs branching fraction, 
the projected reaches for the minimal scalar-portal benchmark with ${\rm BR}(H\to SS)=0$ can be directly applied.
A subset of the viable parameter points, including BP A, lies within the projected sensitivity regions of these experiments. 
In contrast, BP B and the majority of points with ultraweak Higgs--scalar mixing remain beyond the projected reaches considered here. Our results demonstrate that the light-mediator $O(N)$ scalar DM scenario can simultaneously accommodate 
the observed relic abundance, current collider and direct-detection bounds, 
phenomenologically relevant velocity-dependent self-interactions, and, 
for a subset of the viable parameter space, electroweak vacuum stability up to the Planck scale, 
while providing testable targets for future intensity-frontier and long-lived-particle searches.

%%%%%%%%%%%%%%%%%%%%%%%%%%%%%%%%%%%%%%%%%%%%%%%%%%%%%%%%%%%%%
\section*{Acknowledgments}

We acknowledge the use of GPT-5.6 (OpenAI) and Claude Sonnet 5 (Anthropic) 
for assistance in developing and debugging the numerical codes used in this work. 
All AI-assisted code and numerical results were independently checked and verified by the authors, 
who take full responsibility for the scientific content presented here. 
This work was supported in part by Basic Science Research Program 
through the National Research Foundation of Korea (NRF) funded by the Ministry of Science and ICT 
under the Grants Nos. RS-2025-24222969 (J.~L. and S.-h.~N.) and
RS-2025-24533579 (U-R.~K. and S.-h.~N.).

% =========================================================
% References
% =========================================================

\end{document}